\documentclass[aps,twocolumn,pra,showpacs]{revtex4}
\usepackage[dvips]{graphicx}
\usepackage{dcolumn}
\usepackage{bm}

\begin{document}

\author{Jakub S. Prauzner-Bechcicki$^1$, Krzysztof Sacha$^1$, 
and Bruno Eckhardt$^2$}
\affiliation{$^1$Instytut Fizyki imienia Mariana Smoluchowskiego,
  Uniwersytet Jagiello\'nski,
 ulica Reymonta 4, PL-30-059 Krak\'ow, Poland \\
$^2$Fachbereich Physik, Philipps-Universit\" at Marburg, D-35032
Marburg, Germany}

\title{Non-sequential double ionization of molecules in a strong laser field}

\date{\today}

\begin{abstract}
We consider the final stage of double ionization of $\rm O_2$ molecules by 
short linearly polarized laser pulses. The saddles of the effective 
adiabatic potential energy close to which simultaneous escape of electrons 
from a molecule takes place are identified. The analysis 
of the saddles and numerical simulations of the ionization indicate that 
to observe clear signatures of simultaneous electron escape in double ionization of 
$\rm O_2$ molecules stronger and much shorter 
laser pulses than those used in the recent experiment [E. Eremina, {\it et al}, 
Phys. Rev. Lett. {\bf 92},  173001  (2004)] should be applied.
\end{abstract}

\maketitle

\section{Introduction}
Experimental studies of 
a non-sequential double (or multiple) ionization
is possible thanks the use of high intensity ultrashort-pulse lasers~\cite{silap93,silap00}.
On the other hand theoretical description of such multi-electron dynamics is far from complete and
suggests that interactions between electrons become important in considerations~\cite{silap93,silap00,schafer93,yang93}. 
In this paper we present description of the non-sequential double ionization of molecules based on a classical model for electrons
in a combined Coulomb and external field following the approach developed
in~\cite{physica_scripta,eckhardt01pra1,eckhardt01pra2,eckhardt01epl,eckhardt03jpb} 
for the multiple 
ionization of atoms.

Recent experimental investigations aimed on the non-sequential double ionization of diatomic
molecules~\cite{cornaggia98,guo98,guo00,eremina04} 
showed that there are
differences between molecular species. For instance,
in the case of $\rm N_2$ it seems that electrons escape
with similar momenta along field polarization axis more often than in the case
of $\rm O_2$ \cite{eremina04}. Such a correlation in momenta of escaping electrons is viewed as
characteristic feature of the non-sequential double ionization. 
In our previous publication \cite{prauzner04} we have analyzed non-sequential double ionization of molecules 
and presented numerical results mostly for double ionization of $\rm N_2$ molecules. 
Here we complete the analysis by investigating how the distributions change  
in the case of $\rm O_2$ molecules for different parameters of the system.

\section{Results}
In our considerations we assume the re-scattering scenario~\cite{corkum93,kulander93}. Namely, in the
first step one electron tunnels out through
the Stark saddle and then is returned back to the nucleus. 
And so at the expense of the energy brought back by
the returning electron a highly excited state of a molecule is formed.
 Finally, such a highly excited compound
state decays in several ways through a single, double or multiple ionization. We start our analysis
after the formation of the excited complex that is to say we assume that we have an initial state of
two highly excited electrons close to the molecular core in the presence of linearly polarized laser field.
For short laser pulses molecules have not 
enough time to change their orientation~\cite{eremina04,miyazaki04} and thus we presume that the motion of the molecular core is
frozen. Hence, the Hamiltonian reads (in atomic units, which are used throughout):
 \begin{equation}
 H=\frac{{\bf p}_1^2+{\bf p}_2^2}{2}+V,
 \end{equation}
 where the potential,
 \begin{eqnarray}\label{pot}
 V&=& -\sum_{i=1}^2
 \left[ \frac{1}{\sqrt{\left(x_i+\frac{d}{2}\sin \theta\right)^2+y_i^2+\left(z_i+\frac{d}{2}\cos \theta\right)^2}}\right. \cr &&
 \left. +\frac{1}{\sqrt{\left(x_i-\frac{d}{2}\sin \theta\right)^2+y_i^2+\left(z_i-\frac{d}{2}\cos \theta\right)^2}}\right] \cr &&
 +\frac{1}{|{\bf r}_1-{\bf r}_2|}+(z_1+z_2)F(t).
 \end{eqnarray}
 The potential consists of the potential energies associated with interactions of the electrons 
 with the nuclei (the entire structure 
 of the molecular core is approximated 
 by two positively charge nuclei), with each other and with the external field.
Origin of coordinate system is placed in the center of mass of
the nuclei and without loss of generality it is assumed that the molecule lies in the $x$-$z$-plane.
Then $x_i,\ y_i$ and $z_i$ indicate position 
 of the electrons, $d$, the distance between nuclei, 
and $\theta$, the angle between the
molecular axis and the $z$-axis (the polarization axis), respectively. 
 The electric field strength $F(t)$ has an oscillatory component times 
 the envelope from the pulse,
 namely:
 \begin{equation}
 F(t)=F\cos(\omega t + \phi)\sin^2(\pi t/T_d),
 \end{equation} 
with $F$, $\omega$, $\phi$ and $T_d$ the peak amplitude, frequency, initial phase of the field
 and pulse duration, respectively, 
\begin{equation}
T_d=n\frac{2\pi}{\omega}, 
\end{equation}
where $n$ is number of cycles in the pulse.

Within this model the only difference between different 
diatomic species lies in the distance, $d$, between the nuclei. In the present publication 
we restrict ourselves to the oxygen molecule, hence $d=2.28$~a.u. in the following.

To identify channels for simultaneous electron escape
we use an adiabatic approximation, keeping the field fixed, for the  reason that the classical 
motion of the electrons is fast compared to the field oscillations (later, in the numerical simulations, 
we use the potential (\ref{pot}) without any adiabatic assumption). The channels for the non-sequential ionization 
correspond to saddles of the potential (\ref{pot}). That is, when electrons pass close to a saddle they may 
leave a molecule simultaneously because one of unstable directions of a saddle allows both electrons to increase 
distance from a molecule. 
\begin{figure}[ht]
\includegraphics[height=0.4\textwidth,angle=270]{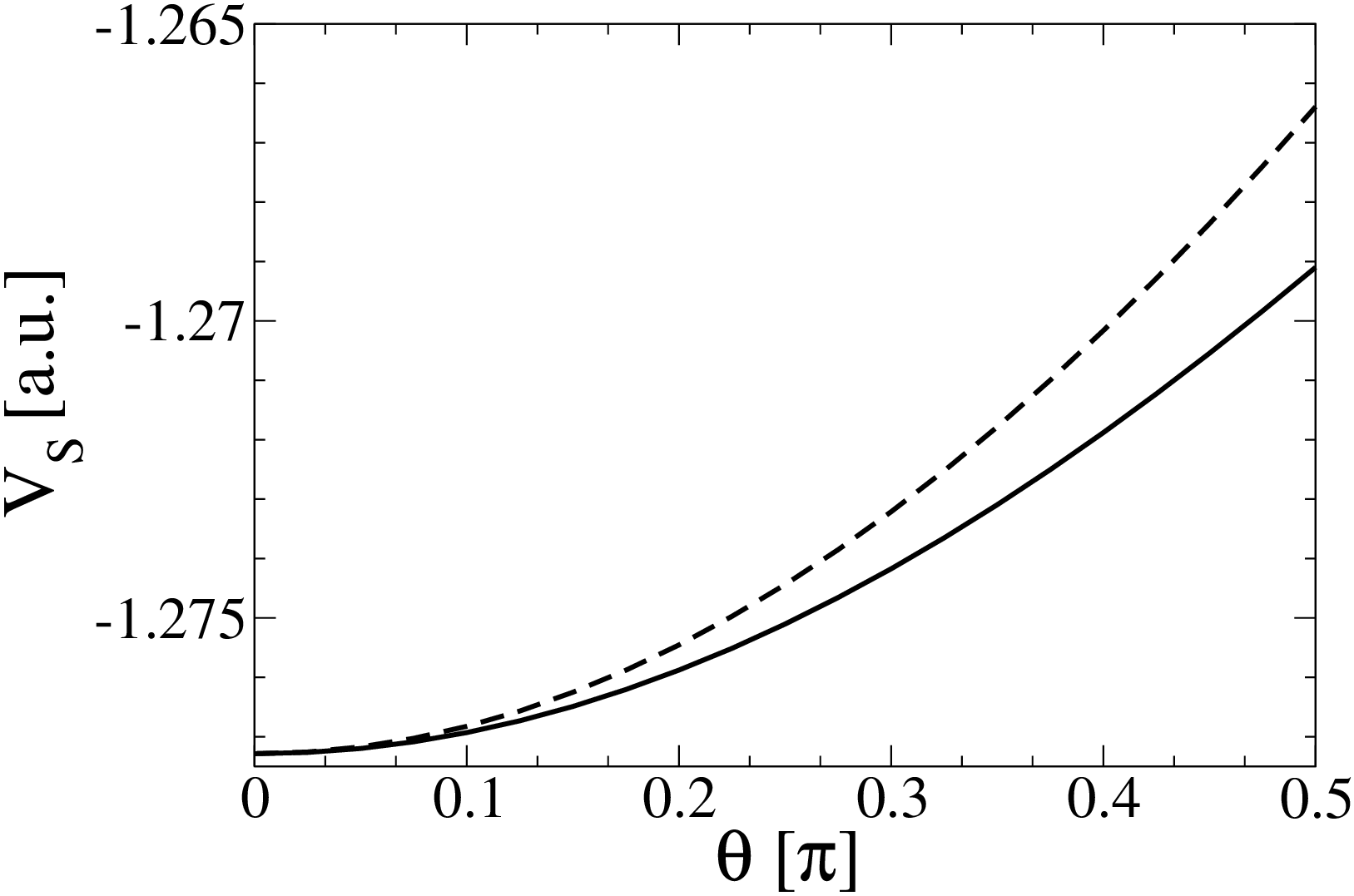}
\includegraphics[height=0.4\textwidth,angle=270]{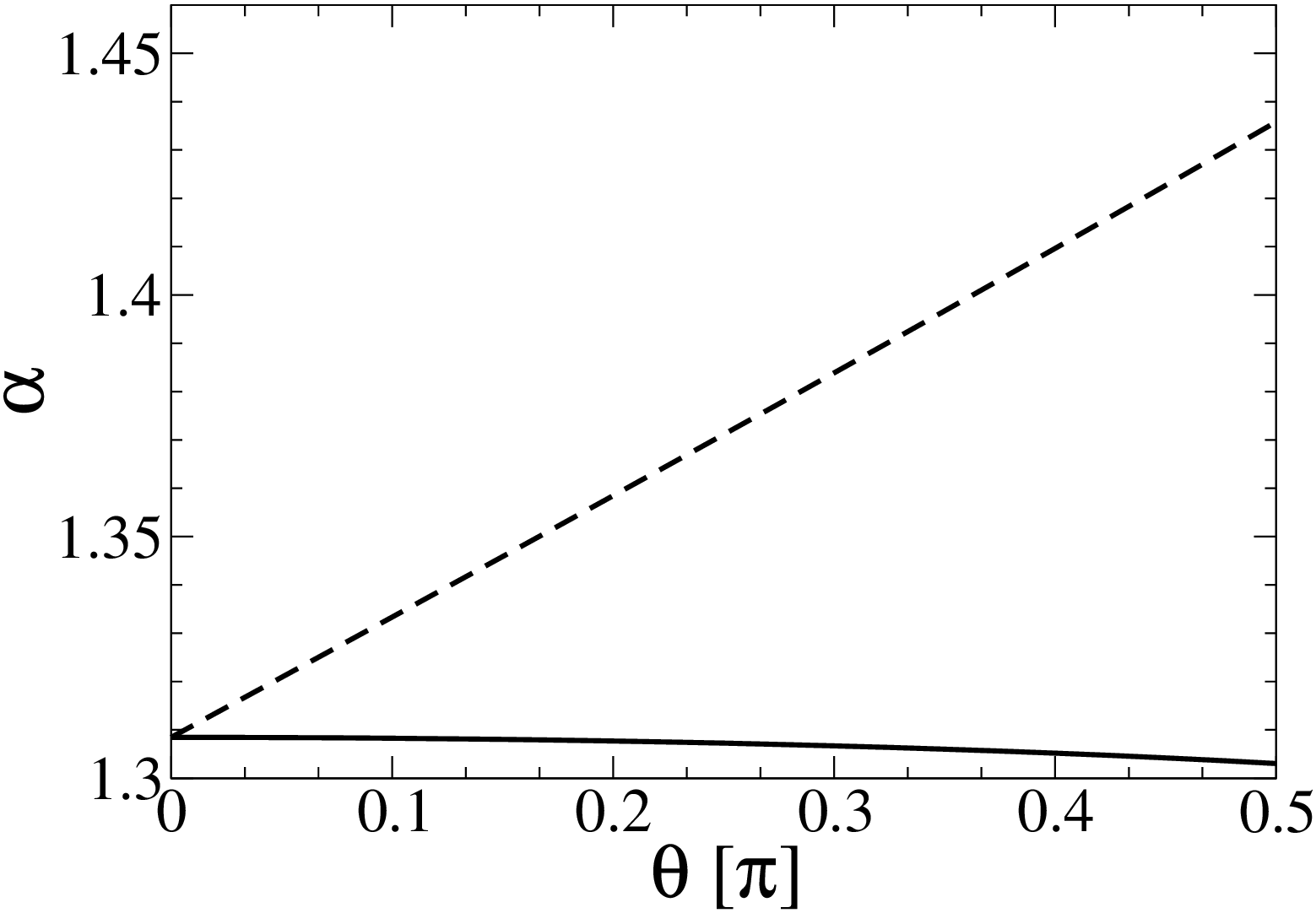}
\caption[]{Energy of the saddle (top panel) and cross section exponent
(bottom panel) as a function of orientation angle, $\theta$, for
O$_2$ molecule ($d=2.28$ a.u.).
Broken line corresponds to the saddle in the $yz$-plane,
whereas solid line corresponds to the saddle in the $xz$-plane. 
The field strength is $F(t)=0.075$ a.u.
\label{angle}}
\end{figure}

For a general orientation of a molecule with respect to the field polarization axis we have found 
two saddles for electron escape. The first is situated in a plane defined by the molecular and field 
axes while the other in the perpendicular plane. 
A few stable and several unstable directions are revealed by the local stability analysis of the
saddles. 
One unstable direction corresponds to the non-sequential double ionization,
 while the other unstable directions reflect interactions that will push electrons away from the 
non-sequential double ionization path leading to the single ionization.

The Lyapunov exponents that characterize the different
directions allow one to obtain the cross section behavior for non-sequential double ionization events.
Similarly to the double ionization without a field, analyzed many
years ago by Wannier~\cite{wannier53,peterkop71,rau84,rost98}, the
competition between the various unstable directions gives rise to an algebraic variation
of the cross section with energy close to the threshold, namely,
\begin{equation}
\sigma (E) \propto (E-V_S)^{\alpha},
\end{equation}
where $V_S$ is the saddle energy and the exponent contains the Lyapunov exponents,
\begin{equation}
\alpha=\frac{\sum_i \lambda_i}{\lambda_r};
\end{equation}
$\lambda_r$ is the Lyapunov exponent of the unstable direction corresponding to 
the non-sequential double ionization path, and $\lambda_i$ are the Lyapunov exponents
of all other unstable directions of the saddle~\cite{rost01phe,eckhardt01epl}.

\begin{figure*}[ht]
\includegraphics[height=0.68\textwidth,width=\textwidth]{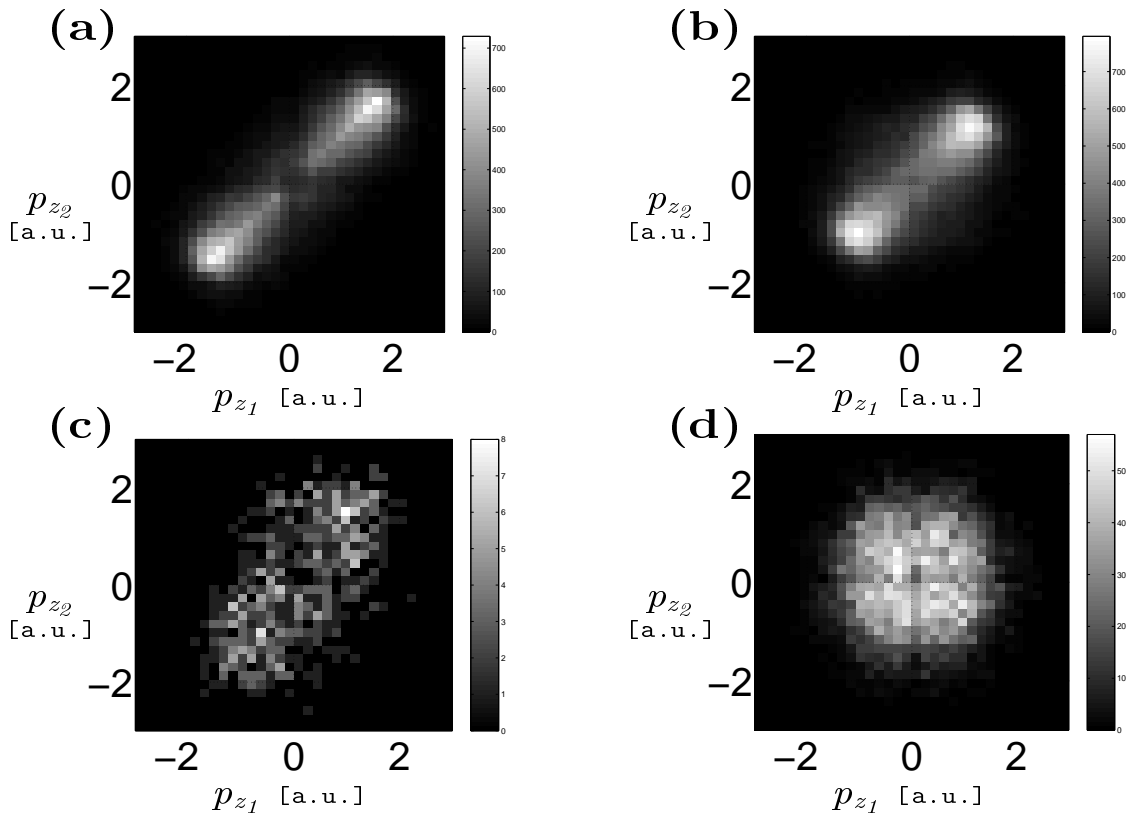}
\caption[]{Final distribution of the parallel electron momenta for different initial energy, 
$E$ and different pulse length (i.e. number of cycles in the pulse, $n$) 
in the double ionization of the $\rm O_2$ molecule\label{theta_zero}. 
The field strength is $F=0.075$~a.u., the frequency $\omega=0.057$ a.u., 
the internuclear distance $d=2.28$ a.u. and the orientation parallel to the field,
$\theta=0$. The plots differ in initial energy $E$ and the number of 
field cycles $n$: (a) $E=-0.05$~a.u., $n=2$; 
(b) $E=-0.05$~a.u., $n=26$; 
(c) $E=-0.9$~a.u., $n=2$ and 
(d) $E=-0.9$~a.u., $n=26$.}
\end{figure*}

In Fig.~\ref{angle} variations of the parameters of the saddles with the orientation 
$\theta$ for the internuclear distance corresponding
to $\rm O_2$ molecule are presented. Fig.~\ref{angle} shows that
the energies and exponents do not change very much with the orientation of the molecule 
(especially in the case of the saddle located in a plane defined by the field and molecular axes).
Thus, we may conclude that from the point of view of the local analysis 
of the non-sequential ionization, decay of the highly 
excited complex does not depend strongly on the orientation of the molecule with respect to the field polarization 
axis.

Thus far we have analyzed the potential within the adiabatic 
approximation. Now we present results of numerical simulations
of decay of a highly excited two electron 
compound state taking into account full time 
dependence of the potential (\ref{pot}).
Before the double ionization escape both electrons  pass close to the nuclei  where 
they interact strongly with each other and with the nuclei. Thus we may assume that all memory of the previous 
motion (i.e. tunneling of the first electron, its evolution in the combined Coulomb and
laser fields
and the rescattering process) is lost. Then it is plausible to assume that the
 compound state which decays to a doubly charged molecule 
can be classically modeled by  a statistical distribution for two electrons close to the nuclei. 
Detailed discussion on numerical calculations and initial conditions is presented
in our previous publication \cite{prauzner04}.

In Fig.~\ref{theta_zero} distributions of the final electron momenta parallel to the field axis for 
the $\rm O_2$ molecule oriented along the field axis are presented. 
Different initial energy $E$ and different pulse durations are considered. 
For very short laser pulses ($n=2$ cycles) the distributions are 
localized along the diagonals indicating 
that the electrons escape predominately by passing close to the saddles 
analyzed previously. For longer pulses and for energy $E=-0.9$~a.u.
the distribution drastically changes its character. Namely, the first and fourth quadrants 
of the panel become strongly populated implying that 
a number of sequential decays significantly increases. 
The reason for this is that, 
after re-scattering when a highly excited two electron complex is created, 
there are two dominant scenarios within 
the first half cycle of the field: 
a non-sequential double escape or a single ionization. During the next 
cycles, unless the molecule is
already doubly ionized, 
we are left with a singly ionized molecule
which may survive to the end of the pulse or the second electron escapes and 
that corresponds to a sequential double ionization.
As the pulse becomes longer, the number of such outcomes
increases, even to the point of overwhelming the number of direct, non-sequential
double ionization cases. Hence, even though the re-scattering scenario is 
involved in the double ionization process, for sufficiently long pulses
 the momentum distributions will not show signatures of the non-sequential electron escape. 
For the initial energy $E$ much higher than 
the minimal energy of a saddle ($V_S\approx -1.27$~a.u.)
the probability of non-sequential ionization is 
bigger and even if the pulse duration is quite 
long (e.g. $n=26$ in panel (b) of Fig.~\ref{theta_zero}) 
the signatures of non-sequential process remain. 

\begin{figure}[t]
\includegraphics[height=0.78\textwidth,width=0.4\textwidth]{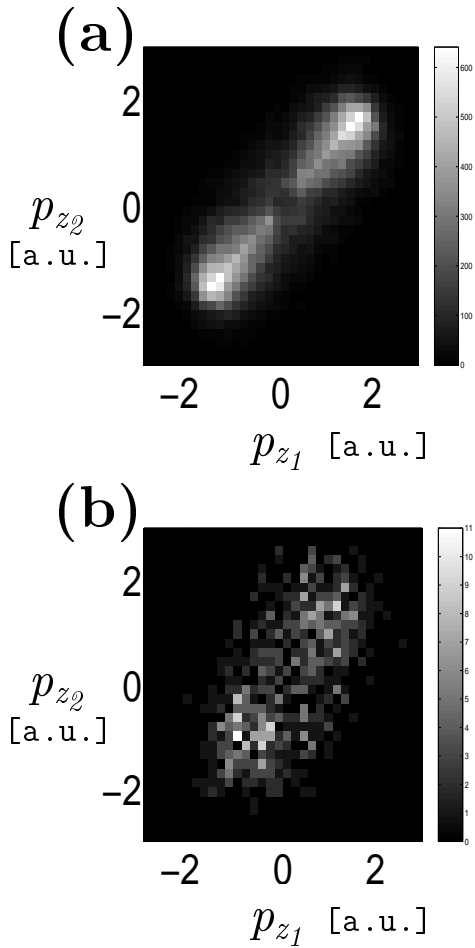}
\caption[]{Final distribution of the parallel electron momenta for the $\rm O_2$ molecule 
at non-zero angle to the field axis, i.e. $\theta=\pi/4$, and for different initial energy $E$. 
$F=0.075$~a.u., $\omega=0.057$~a.u., $d=2.28$~a.u. and $n=2$. The energies are (a) $E=-0.05$~a.u. 
and (b) $E=-0.9$~a.u.}
\label{dangle}
\end{figure}

Alteration in the orientation of the molecule does not lead to significant changes
 in the momentum distributions as one can see by comparing 
Fig.~\ref{dangle} and Fig.~\ref{theta_zero}. For the field amplitude 
used in the simulations the saddles for non-sequential 
process are far away from the nuclei and the 
positions and other parameters of the saddles change only slightly 
with the vary of orientation $\theta$ --- compare Fig.~\ref{angle}. 

Figs.~\ref{theta_zero}-\ref{dangle} should be compared with the corresponding figures for nitrogen
molecule presented in Ref.~\cite{prauzner04} --- there we chose 
energies $E=-0.3$~a.u. and $E=-0.6$~a.u. (also the field peak amplitude was slightly
smaller, i.e. $F=0.07$~a.u.). Comparing the data we may conclude that high initial energy 
leads to much clearer signatures of simultaneous electron escape (even for long pulses) 
than low energy. Initial energy is determined in the rescattering process, thus to enhance
signatures of the simultaneous escape one should apply stronger external field.
 Increasing field amplitude is, however, dangerous because for very strong laser pulses
  sequential ionization (without rescattering) becomes dominant.
   Compromise can be reached by increase of the amplitude with simultaneous reduction of the  
pulse duration --- very short laser pulses and the field stronger than that used in Ref.~\cite{eremina04} 
should allow for clear observation of simultaneous electron escape in double ionization of 
$\rm O_2$ molecules.

Our analysis is restricted to the final stage of the non-sequential double ionization 
of $\rm O_2$ molecules, where we choose initial conditions microcanonically.
 Early stages of the excitation
process and the nature of the compound state
 (i.e. the symmetry of the valence orbital may affect the excitation process)
 before the final decay towards double ionization may have influence on the initial
  conditions. That, in turn, may modify the final parallel momenta distributions. This aspect needs
  further investigation.

\section{Summary}
\label{summary}

We have performed a purely classical analysis of the final stage 
of the non-sequential 
double ionization of $\rm O_2$ molecules in the strong laser field where a molecular core is approximated by two 
positively charged nuclei and the re-scattering scenario is assumed. 
Our analysis, 
together with the previous studies 
devoted mostly to double ionization of $\rm N_2$ molecules \cite{prauzner04}, 
allow us to draw following conclusions: 
i)~From the point of view of 
classical analysis, within the considered model,
there is no difference between nitrogen and oxygen molecules in the sense 
that both of them can show signatures
of simultaneous double escape. ii)~Orientation of the molecule with respect 
to the field axis 
does not influence significantly the final momentum distribution for the 
initial energy range considered in
the model. iii)~Shorter and stronger laser pulses
should enhance signature of simultaneous electron escape that will 
be visible in the momentum distributions as a more pronounced symmetrical escape of the electrons.

\section{Acknowledgments}
This work  was partly supported by the Polish Ministry of 
 Scientific Research Grants: 
 PBZ-MIN-008/P03/2003 (JPB), 1P03B09327 (KS) and 
 by the Deutsche Forschungsgemeinschaft.

\end{document}